\documentclass{article}
\usepackage[english]{babel}
\usepackage{amsmath,mathrsfs,amssymb}   

\usepackage[latin1]{inputenc}

\usepackage{natbib}

\usepackage[dvips]{graphicx}
\usepackage[dvips]{color}
\graphicspath{{Figs/}}

\usepackage{bm,bbm}
\usepackage{psfrag}

\newcommand{\ba}{\begin{eqnarray}}
\newcommand{\ea}{\end{eqnarray}}
\def\H{H}
\def\K{K}

\def\pt{{\partial}}
\def\ev{{\bf e}}
\def\nv{{\bf n}}
\def\nvs{{\bf n}_{s}}
\def\tavs{\tav_{s}}
\def\mv{{\bf m}}

\def\phiv{\boldsymbol{\varphi}}
\def\bnus{\boldsymbol{\nu}_{\hspace{-0.5mm} s}}
\def\cn{c_{\nvs}}
\def\ct{c_{\tavs}}
\def\tv{{\bf t}}

\def\tv{\boldsymbol{\tau}}
\def\uv{{\bf u}}
\def\vv{{\bf v}}
\def\kv{{\bf k}}

\def\Av{{\bf A}}
\def\kg{\kappa_{\nvs }}
\def\Bv{{\bf B}}

\def\Iv{{\bf I}}

\def\Lv{{\bf L}}
\def\Mv{{\bf M}}
\def\Mvs{\overline{\bf M}}
\def\Ms{\overline{M}}
\def\Qvs{\overline{\bf{Q}}}

\def\kk{\kappa_{\tavs }}
\def\sph{{\mathbb S}}
\def\re{{\mathbb R}}
\def\tg{\tau_{\nvs }}
\def\cu{c_{1s}}
\def\cc{c_{2s}}
\def\tav{\mathbf{t}}

\def\Pv{{\bf P}}
\def\Qv{{\bf Q}}

\def\eps{ \varepsilon}
\def\dd{{\rm d}}

\def\tr{{\rm tr}}

\def\grad{\nabla}
\def\grads{\nabla\hspace{-1mm}_s}
\def\MS{\mathbf{M}_s}
\def\rots{\mathrm{curl}_s}
\def\rot{\mathrm{curl}}

\def\dv{{\rm div}}
\def\dvs{\dv \hspace{-0.5mm}_s}
\def\esse{{\cal{S}}}

\newcommand{\ot}{\otimes}

\newcommand{\bnu}{ \mbox{ \hspace{-1mm}\boldmath $\nu$ \hspace{-1mm}}}
\newcommand{\ricci}{ \mbox{ \hspace{-1mm}\boldmath $ \epsilon$ \hspace{-1mm}}}

\newcommand{\bom}{ \mbox{ \hspace{-1mm}\boldmath $\omega$ \hspace{-1mm}}}

\newtheorem{proposition}{Proposition}
\begin{document}

\title{\bf Surface free energies for nematic shells}
\author{Gaetano Napoli$^1$  \and Luigi Vergori$^2$}
\date{ \small $^1$Dipartimento di Ingegneria dell'Innovazione,
  Universit\`a del Salento, via per Monteroni, Edificio ``Corpo O'',
  73100 Lecce, Italy \vspace{5mm}\\
  $^2$Dipartimento di Matematica, Universit\`a del Salento, Strada Prov. Lecce-Arnesano, 73100 Lecce, Italy}

\maketitle
\begin{abstract}
{We propose a continuum model to describe the molecular alignment  in thin nematic shells. By contrast with previous accounts, the two-dimensional free energy, aimed at describing the physics of thin films of nematics deposited on curved substrates, is not postulated but it is deduced from the conventional three-dimensional theories of nematic liquid crystals.  Both the director and the order-tensor theories are taken into account. The so-obtained surface energies exhibit extra terms compared to earlier models.  These terms reflect the coupling of the geometry of the shell with the nematic order parameters.  As expected, the shape of the shell plays a key role in the equilibrium configurations of nematics coating  it. }
\end{abstract}

\vspace{5mm}

\vspace{5mm}

{\bf keywords:} Nematic shells, nematic membranes, two-dimensional nematic order

\section{Introduction}
Nematic liquid crystals are aggregates of rod-like molecules that tend to align parallel to each other along a given direction \citep {degennes}.  Due to their  easy response to externally applied electric, magnetic, optical and surface fields, liquid crystals are of greatest potential for scientific and technological applications. Currently, there is an increasing interest  in soft matter physics on small spherical colloidal particles or droplets coated with a thin layer of nematic liquid crystal  \citep{lopez:2011,lopez-leon:2011prl}. The hope is  to build {\it mesoatoms}  with controllable valence   \citep{nelson:2002}.  We refer to these coating layers as {\it nematic shells}.  

When nematic liquid crystals are constrained to a curved surface, the geometry induces a distortion in the molecular orientation. The possibility to have an in-plane order rather than a spatial distribution of the molecules, depends on the shell thickness \citep{vitelli:2006, Fernandez:2007, lopez:2011}.  In ultra-thin shells, the interaction with the colloid surface enforces a sort of degenerate anchoring, {\it i.e.} the tendency of the molecules to align along any direction parallel to the surface. Thus, unavoidably defects arise when nematic order is established on a surface with the topology of the sphere.  The number of defects is a consequence of the  Poincar\'e-Hopf theorem, that states that any configuration must have a total topological charge equal to the Euler-Poincar\'e characteristic of the surface. For instance on a sphere, whose characteristic is $+2$, we can have two diametrically opposite $+1$ defects or  four $+1/2$ defects located at the vertices of a tetrahedron \citep{vitelli:2006}. This tetrahedral defect structure is of great interest in material science because defects  regions  can be functionalized to serve as bonds \citep{nelson:2002}. This  could lead to tetravalent mesoatoms with $sp^3$-like directional bonding  like carbon.  Theoretical studies have emphasized the possibility to control the location of the defects, and hence the valence of mesoatoms,  by varying the shell geometry \citep{kralj:2011} or by tuning the elastic constants of the nematic \citep{bates:2008,  shin:2008}.
 
 Most theoretical studies on nematic liquid crystals are framed within the classical {\it director theory} (see, for instance, \citep{degennes,virga}). In this setting, the local properties of the nematic liquid crystal are described through a unit vector, the director, parallel to the local molecular direction. The equilibrium configurations of the nematics minimize the Frank's free energy, with respect to all configurations that satisfy  the boundary conditions. However, the director description of a nematic configuration misses a relevant information at the mesoscopic level:  the dispersion of the molecules around the director. The {\it order-tensor theory}, put forward by de Gennes (see \cite{degennes, virga}), overcomes  this gap by introducing a richer kinematic description. Within this theory the free energy to minimize is the Landau-de Gennes free energy.  
 
Theories for two-dimensional nematic order  have been proposed in both director and order-tensor schemes \citep{straley:1971, helfrich:1988, lubensky:1993, selinger:2001, biscari:2006,kralj:2011} and use free energies derived by symmetry arguments or mesoscopic properties. By contrast,  our approach derives the surface free energy for thin films as limiting cases of the well established  three-dimensional theories of nematic liquid crystals.   The main concern  is how  classical theories (Frank and  Landau-de Gennes theories)  reduce when the nematic molecules, confined within a thin region,  align in the direction parallel to the underlying surface. A prominent role is played by the ratio between the thickness of the shell, denoted by $h$, and    the minimum radius of curvature of the entire shell, denoted by $\ell$. In fact, the  surface versions of Frank  and  Landau-de Gennes free energies can be deduced from the three-dimensional models under the assumption of smallness of the ratio $h/\ell$. 

Conversely to existing models, we find that in the two-dimensional directory theory the twist term does not vanish.  Actually, it expresses the tendency of the molecular axis to align with the principal directions. Moreover, our analysis provides a coherent way to obtain the two-dimensional order-tensor  theory.  As a result, we retrieve the quadrupolar coupling between the two-dimensional order tensor and the curvature tensor already obtained by \cite{biscari:2006} using  mesoscopic arguments.  

The paper is organized as follows. In Section $2$, we introduce  the mathematical notation and terminology. Sections $3$ and $4$, are devoted to obtain  surface free energies  from Frank  and Landau-de Gennes theories, respectively. Mathematical topics  employed in these sections and some  details of the calculations are reported in the Appendixes.   Finally, we draw our conclusive remarks in Section $5$.     
\section{Geometrical preliminaries}
We first introduce the  terminology and establish some preliminary notations. First of all, three-dimensional vectors are denoted by lower-case boldface letters, whereas second order tensors are denoted by upper-case boldface letters. The scalar, vector and tensor products between two vectors $\uv$ and $\vv$ are indicated by $\uv \cdot \vv$, $\uv \times \vv$ and $\uv \otimes \vv$, respectively. In cartesian components, $\uv \cdot \vv = u_i v_i$, $(\uv \times \vv)_i =  \epsilon_{ijk} u_j v_k$, $(\uv \otimes \vv)_{ij} = u_i v_j$,  where summation is understood over repeated indices, and the third order tensor $\ricci= \epsilon_{ijk}$ is the Ricci alternator.  The composition between two second-order tensors $\Av$ and $\Bv$ is the tensor $\mathbf{C}=\Av\Bv$ with components $C_{ij} = A_{ih} B_{hj}$, whereas  the composition between a second order tensor $\mathbf{A}$ and a vector $\mathbf{u}$ gives the vector $\mathbf{v}=\Av\uv$ with components $v_i=A_{ij}u_j$. Finally, the scalar product between $\Av$ and $\Bv$ is the scalar $A_{ij}B_{ij}$.

Let us assume that the nematic shell occupies a thin region $V$ of thickness $h$ around a regular compact surface $\esse$. Let $\bnus$ be the normal unit vector field to $\esse$. We parameterize points in the bulk through a coordinate set $(u,v,\xi)$ such that
\ba \label{para}
p(u,v,\xi) =  p_S(u,v) + \xi \bnus (u,v),
\ea
where $p_S$ is the normal projection of $p$ onto $\esse$, and  {$|\xi|$}, with $\xi\in[-h/2,h/2]$, is the distance of $p$ from the same surface. Such a coordinate set is well defined in a finite neighborhood of $\esse$.  More precisely,   we introduce the principal curvatures $c_{1s}(p_\esse)$ and $c_{2s}(p_\esse)$ of $\esse$ at point $p_\esse$, and  assume
\ba \label{hl}
h \ll \min_{p_\esse\in\esse}\left(\max\{|c_{1s}(p_\esse)|, |c_{2s}(p_\esse)|\}\right)^{-1} = \ell.
\ea

For every fixed $\xi\in[-h/2,h/2]$, equation (\ref{para}) defines a parallel surface $\esse_{\xi}=\{p_\esse+\xi\bnus(p_\esse):p_\esse\in\esse\}$ at distance $|\xi|$ from $\esse$ with the vector field $\bnu:p\in\esse_\xi\mapsto \bnus(p_\esse)$ as unit normal vector field. In such a way, the unit vector field $\bnu$ is defined on the entire region $V$. The second-order tensor {$\grad \bnu$} is symmetric. Its eigenvectors  are  $\bnu$ (with a null eigenvalue)  and the unit vector fields
 $$\ev_i(p) = \ev_{is}(p_\esse)\quad (i=1,2),$$
where $\ev_{1 s}$ and $\ev_{2 s}$ represent the tangent principal directions fields  on $\esse$.    The spatial gradient for each  eigenvector is 
\ba \label{guno}
\nabla\bnu=-\frac{\cu}{1-\xi \cu}\ev_1\ot\ev_1-\frac{\cc}{1-\xi \cc}\ev_2\ot\ev_2,
\ea
\ba \label{gdue}
\nabla \ev_1= \kappa_1(\xi) \ev_2\ot\ev_1+ \kappa_2(\xi)\ev_2\ot\ev_2+\frac{\cu}{1-\xi \cu}\bnu\ot\ev_1,
\ea
\ba \label{gtre}
\nabla \ev_2=- \kappa_1(\xi)\ev_1\ot\ev_1- \kappa_2(\xi) \ev_1\ot\ev_2+\frac{\cc}{1-\xi \cc}\bnu\ot\ev_2,
\ea
where the functions $\kappa_1$ and $\kappa_2$ are given in Appendix A. We refer the reader to the book of \cite{docarmo} for a more comprehensive treatise of the geometry of surfaces. 

Let $\Phi$ be a smooth field defined on $\esse$. Assume $\Phi$ scalar, vector or tensor valued. Then the surface gradient of $\Phi$ is defined (see \cite{gurtin:1975}) as
$$
\grads \Phi= (\grad \Phi) \Pv,
$$
where  $\Pv=\mathbf{I}-\bnus\ot\bnus$ represents the projection onto the tangent plane of $\esse$. The trace of the surface gradient of a  vector field $\mathbf{u}$ defines the surface divergence of $\mathbf{u}$: $\dvs\mathbf{u}=\mathrm{tr}\grads\mathbf{u}=\grads\mathbf{u}\cdot \Pv$, that is a scalar field. On the other hand, the surface curl of $\mathbf{u}$ is defined as twice the skew-simmetric part of the surface gradient:
$$
\rots\mathbf{u}=-\ricci\grads\mathbf{u},
$$
where $\ricci$ denotes the Ricci alternator.

Let   $\nv$ denote a unit vector field defined on $V$ such that  $\nv(p)= \nv(p_S)$  and 
$
\nv\cdot\bnu = 0
$
 at each point $p$.   Next, by introducing the {\it conormal} unit vector field $\tav = \bnu \times \nv$, we  write the spatial gradient of $\nv$ (see Appendix A for calculation details):
\begin{align}\label{gradn}
 \nabla\nv&=\iota^{-1}\Big\{ \left[\kg-\xi\bnus\cdot\rots(\Lv\nvs)\right] \tav\ot\nv+[\kk-\xi\bnus\cdot\rots(\Lv\tavs)]\tav\ot\tav
\nonumber \\
& +(\cn-\xi\K)\bnu\ot\nv- \tg\bnu\ot\tav\Big\},
\end{align}  
where $\nvs$ and $\tavs$ represent the restrictions of $\nv$ and $\tav$ on $\esse$, respectively. The tensor $\Lv=-\grads\bnus$ represents the extrinsic curvature tensor of $\esse$. Its trace gives twice the mean curvature $H$, while its determinant gives the Gaussian curvature $K$.  The quantities $\cn = \nvs \cdot \Lv \nvs$, $\tg=-\tavs \cdot \Lv \nvs $ represent the normal curvature and the  geodesic torsion of the flux lines of $\nvs$ on $\esse$, respectively.  The latter is zero whenever $\nvs$ is a principal direction. The quantities $\kg$ and $\kk$ denote the  geodesic curvature of the flux lines of $\nvs$ and $\tavs$ on $\esse$, respectively \citep{rosso:2003, tu:2004}.  The geodesic curvature $\kg$  (respectively, $\kk$) measures the deviance of the flux lines of $\nvs$ (respectively, $\tavs$) from following a geodesic on $\esse$.   Finally, we have set $\iota = 1-2\H\xi+\kg\ xi^2$.

The divergence  and the $\rot $ of $\nv$ are the trace of $\grad \nv$ and the axial-vector associated with  twice  the skew-symmetric part of $\grad \nv$, respectively. Thus, from (\ref{gradn}) it follows that
\ba
 \dv\nv=\iota^{-1}\left[\kk-\xi\bnus\cdot\rots(\Lv\tavs)\right],
\label{divergenza}
\ea
\begin{align}\label{rotore}
 \rot\nv= \iota^{-1}\big\{-\tg \:\nv- (\cn-\xi\K) \:\tav+[\kg-\xi\bnus\cdot\rots(\Lv\nvs)] \bnu\big\}.
\end{align}
We observe  that the normal curvatures, the geodesic torsion, the geodesic curvatures and the surface gradients introduced above are quantities related to the surface $\esse$ and, therefore, they do not depend on $\xi$. Instead, although $\nv$ has been supposed constant along normal directions within the thickness, its spatial gradient depends on $\xi$.

Finally, since $\kg= \tavs \cdot (\grads\nvs)\nvs$ and $\kk=\tavs \cdot  (\grads\nvs)\tavs$ (see \cite{rosso:2003}), the surface gradient of $\nvs$ is  given by 
\ba
\grads\nvs = \kg\ tavs\ot\nvs+ \kk \tavs\ot\tavs + \cn\bnus\ot\nvs-\tau_{\nvs} \bnus\ot\tavs, \nonumber
\ea
 and consequently
\ba\label{dvsrots}
 \dvs \nvs = \kk,  \qquad   \rots \nvs = -\tau_{\nvs} \nvs-\cn\tavs+\kappa_{\nv} \bnus.
\ea
Unlike flat surfaces,  the surface curl of $\nvs$ possesses nonvanishing  in-plane components.

\section{Two-dimensional director theory}
The classical elastic continuum theory is based on the pioneering works of Oseen,  Zocher and Frank published between the thirties and the fifties of last century. We refer the reader to the book of \cite{virga}  for a detailed mathematical treatment.  The average alignment of the molecules is represented by a unit vector $\nv$, called the director, where $\nv$ is physically equivalent to $-\nv$.  The expression for the elastic energy density (per unit of volume) associated with the director distortion consists of four terms
\begin{align}
2w_{OZF} & = K_1 (\dv \nv)^2 + {K_2}(\nv \cdot \rot \nv)^2 + {K_3}|\nv \times \rot \nv|^2 \nonumber \\ &+ (K_2 + K_{24})\dv[(\nabla \nv)\nv - (\dv \nv)\nv]
\label{OZF}
\end{align}
where the	constants	$K_1$, $K_2$, $K_3$, and $K_{24}$ are called the splay, twist, bend, and saddle-splay moduli, respectively. To ensure a stable undistorted configuration of a nematic liquid crystal in the absence of external fields or confinements, the three moduli $K_i (i = 1, 2, 3)$ must be non-negative, whereas the elastic saddle-splay constant must obey Ericksen's inequalities [6]:
$$ 
|K_{24}|\leq K_2, \qquad	K_2 +K_{24}\leq 2 K_1.
$$
In the absence of external actions, the equilibrium configurations are stationary points of the total energy
\ba
W = \int_{V} w_{OZF}(\nv, \grad \nv)  \dd V,
\ea
 according to the boundary conditions. These may consist in fixing $\nv$ at the boundary (strong boundary conditions) or in allowing $\nv$ to  rotate freely (free boundary conditions). Intermediate situations, known as {\it weak anchoring} boundary conditions, can be envisaged by including an anchoring energy  that penalizes the deviation of the molecules at the boundary from a given direction. Furthermore, the free energy density  may account for extra terms in order to describe, for instance,  the interaction of the nematic with external electric or magnetic fields. 

Let us introduce the small parameter $\eps=  h/\ell$. 
The smallness of $  \varepsilon$ on the one hand ensures that the parameterization (\ref{para}) is properly defined and on  the other hand, with the aid of Proposition \ref{Frank theorem} (see Appendix B), it allows us to approximate  the OZF free energy  as follows
 \ba \label{OZFshell1}
W_{OZF}\approx W^S_{OZF} =\frac{1}{2} \int_\esse \left[k_1 (\dvs\nvs)^2 +k_2(\nvs\cdot\rots\nvs)^2+k_3|\nvs\times\rots\nvs|^2\right]\dd A, 
\ea
where  $k_i = h K_i$ ($i=1,2,3$). Observe that the saddle-splay term has disappeared in this approximation since $\nv$ has assumed to be  constant throughout the thickness. In fact, from (\ref{gradn}) it follows 
\begin{equation}\label{null ss}
\dv[(\grad\nv)\nv-(\dv\nv)\nv]=\mathrm{tr}(\grad\nv)^2-(\mathrm{tr}\grad\nv)^2=0.
\end{equation} 

Comparing equations (\ref{OZFshell1}) and (\ref{OZF}) we remark that: (i)  $W^S_{OZF}$ involves a surface integral rather than a volume integral, thus we can refer to  $W^S_{OZF}$ as a {\it surface free energy}; (ii) the surface elastic constants $k_i$ are obtained by multiplying $K_i$ and the thickness $h$, and, hence, by virtue of  Ericksen's inequalities, they must be non negative; (iii)  the surface free energy involves surface differential operators instead of  spatial ones.

It is worth mentioning  a peculiarity of curved substrates with respect to planar nematics. Unlike the planar case, the twist term cannot be a priori neglected. Indeed, as it has been already observed,  $\rots \nvs$ is not orthogonal to $\nvs$.  In fact,  by using formulae (\ref{dvsrots}), equations (\ref{OZFshell1}) reduces to
\ba
 W^S_{OZF} = \frac{1}{2}\int_\esse \left[k_1 \kk^2 +k_2 \tg^2+k_3(\cn^2 + \kg^2)\right]\dd A,
\label{OZFshell}
\ea
which shows that the twist free energy density is proportional to  $\tg^2$. The latter vanishes if and only if the flux lines of $\nvs$ lie along  principal directions. Thus, the twist free energy can be disregarded  whenever  spherical shells are concerned \citep{shin:2008} or whenever the director lies along meridians or parallels of an axisymmetric shell \citep{kamien:2009}.  

In  light of (\ref{OZFshell}), we can give the following intuitive interpretation for the shell-nematic interaction. The arrangement of the molecules on a surface is the result of the competition between the splay and the bend  free energies that try to put the flux lines of  $\nvs$ and $\tavs$ along  geodesics of $\esse$, and the twist term that tries to align the flux lines of  $\nvs$ with  the curvature lines of $\esse$. Furthermore, the term proportional to the square of the normal curvature, expresses the tendency of the flux lines of $\nvs$ to align with the principal direction of minimal  curvature.    

From equation (\ref{OZFshell}) it follows that within  the {\it one constant approximation}  $(k_1=k_2=k_3=k)$, the surface OZF free energy  becomes 
 \ba \label{OZFshellone}
  W^S_{OZF}= \frac{k}{2} \int_S |\grads \nvs|^2 \dd A.
 \ea
A key feature of the free energy (\ref{OZFshellone}) is that it differs from the one used in earlier works \citep{straley:1971, vitelli:2006, seifert:2007}. Indeed, by denoting $\alpha$ the angle between the principal direction $\ev_{1s}$ and $\nvs$, equation (\ref{OZFshellone}) reduces to
\ba
 W^S_{OZF}= \frac{k}{2} \int_S (|\grads \alpha - \bom |^2 + \cn^2 + \tv^2) \dd A, \nonumber
\ea
where $\bom$ represents the {\it spin connection field} \citep{nelson:1987, giomi:2009}.  A glance at equation (21a) of \cite{nelson:1987} shows that the terms proportional to $\tg^2$ and $\cn^2$ were neglected.  Clearly, this mismatch stems from the fact that free energy density in (\ref{OZFshellone}) is proportional to the square of the surface gradient of $\nvs$ rather than proportional to the square of covariant derivative of $\nvs$ as it is customary to assume  (see for instance \citep{nelson:1987} or  \citep{seifert:2007}).

\section{Two-dimensional order-tensor  theory}
The director theory describes only states with a single  optical axis. For closed shells whose topology is different from that of a torus, the tangent vector field $\nv$ exhibits singular points, {\it i.e.} regions where the local orientational order of the nematic is undefined.  As a result, the shell often incorporates so-called topological defects. These mathematical singularities can be avoided  by introducing a tensorial-order parameter, that describes defects as those points in which the nematic melts into a liquid phase (isotropic states). Hereinafter we illustrate the geometrical meaning of that order parameter. 

We now recall the order-tensor theory for the usual three-dimensional nematics. Let us suppose that the orientation of a single molecule is represented by a unit vector $\mv$. Microscopic disorder is taken into account by introducing a probability measure	$f_p: \sph^2 \rightarrow \re^+$,	such	that	$f_p(\mv)$ describes the probability that a molecule placed in $p$ is oriented along $\mv$.  The orientation of the molecular axis is described at each point in space by a point of the unit sphere $\sph^2$ (or by a unit vector). Thus, if $\Omega$ is any subset of $\sph^2$, the probability of finding in $p$ one molecule oriented within $\Omega$ is given by
\ba
P\{\Omega\} = \int_\Omega f_p(\mv) \dd \sigma \nonumber,
\ea
where $\sigma$ denotes the area measure on $\sph^2$. Nematics posses a molecular mirror symmetry, {\it i.e.}, the {\it head} and {\it tail} of a molecule can be changed without experiencing any change in the probability distribution. Thus, the probability measure is even,  $f_p(\mv) = f_p(-\mv)$, and the first moment of the distribution $f_p$ is zero.

The second moment of the distribution is the variance tensor  $\Mv = \langle \mv \ot \mv \rangle$, where  the brackets denote averaging with respect to $f_p$.  By definition, $\Mv$ is unit trace symmetric and semidefinite positive.  The spectral decomposition theorem ensures that $\Mv$ can be put in the diagonal form:
\ba
\Mv = \lambda_1 \ev_1 \ot \ev_1 +  \lambda_2 \ev_2 \ot \ev_2 +  \lambda_3 \ev_3 \ot \ev_3, \nonumber
\ea
and, since the eigenvalues of $\Mv$  sum up to one,  its spectrum is bounded by ${\rm sp} (\Mv) \subset [0, 1]$.

Nematics may exhibit three different states: isotropic, uniaxial, and biaxial.  It is customary to define these states by using the {\it order tensor}  {$\Qv = \Mv - \frac{1}{3}\Iv$}.   Thus, we can have:
\begin{itemize}
\item[(\emph{i})]  the eigenvalues of $\Qv$ are equal, which  implies $\Qv _{iso}={\bf 0}$; in this case we label the nematic as {\it isotropic}.
\item[(\emph{ii})] At least two eigenvalues are equal, the nematic is called {\it uniaxial}. Simple algebraic manipulation allows us to write:
\ba
\Qv_{uni} = s\left(\uv \ot \uv - \frac{1}{3}\Iv\right). \nonumber
\ea
The scalar parameter $s\in[-\frac{1}{2},1]$ is called the {\it degree of orientation}, while the unit vector $\uv$ is the optical axis. We retrieve the isotropic phase when $s=0$, while the perfect alignment of the molecules is obtained for $s=1$. The case $s=-\frac{1}{2}$ represents flat isotropic distributions, in the plane orthogonal to $\uv$.   
\item[(\emph{iii})]  When the eigenvalues of the order tensor are all different, the nematic is labeled as {\it biaxial}. 
 Then we can write the general expression for the order tensor
\ba
\Qv_{bia} =   s\left(\uv \ot \uv - \frac{1}{3}\Iv\right)+ \lambda\left(\ev_+ \ot \ev_+   -  \ev_- \ot \ev_- \right), \nonumber
\ea
where $\lambda$ denotes the degree of biaxiality and $s\in[-\frac{1}{2},1]$ as above. The sign of $\lambda$ is unessential, since it only involves an
exchange between $\ev_+$ and $\ev_-$. The degree of biaxiality does always satisfy $|\lambda| \leq \frac{1}{3}(1-s)$.  Even for biaxial nematic, $s=-\frac{1}{2}$ represents flat (non necessarily isotropic) distributions.
\end{itemize}

The free energy comprises two terms: the elastic energy and the Landau-de Gennes potential. Following \cite{longa:1987}, the most general quadratic elastic energy can be written as
\ba\label{elastic energy}
W_{el}(\grad \Qv, \Qv) = \int_V \left[ L_{1} Q_{ij,k} Q_{ij,k}  +  L_{2} Q_{ij,k} Q_{ik,j} + L_{24} (Q_{ij,k}Q_{jk}-Q_{ij}Q_{jk,k})_{,i} \right]  \dd V,
\ea
where $L_1$, $L_2$ and $L_{24}$ are constants. Here a comma denotes a partial derivative with respect to one of the coordinates. This energy expresses the tendency of the molecules to arrange  parallel one  to  each other in a homogeneous state.

The Landau-de Gennes potential, $W_{LdG}$, is a temperature-dependent thermodynamic contribution that takes into account the material tendency to spontaneously arrange in ordered or disordered states. Its density is of the form (see \cite{degennes})
\ba
\label{enldg}
 w_{LdG}(\Qv)=F(A,B,C)+\frac{A}{2} \tr \Qv^2 - \frac{B}{3} \tr \Qv^3 + \frac{C}{4} (\tr \Qv^2)^2,
\ea
where $A=A_0(T-T_c)/T_c$, $A_0$ is a  material-dependent positive constant, $T$ is the absolute temperature and $T_c$ is a characteristic temperature; $B$, $C$ are material-dependent positive constants and $F(A,B,C)$ is a positive constant that accounts for the free energy of the isotropic phase. We observe that $F(A,B,C)$ plays no role in the minimization of the Landau-de Gennes energy density and the stationary points of $w_{LdG}$ correspond to either isotropic tensors or, whenever $B^2-24AC\geq0$,  uniaxial tensors of the form
$$
\Qv_{cr}=\tilde{s}\left(\uv\ot\uv-\frac{1}{3}\Iv\right),
$$
with 
$$
\tilde{s}=\frac{B+\sqrt{B^2-24AC}}{4C},
$$
and $\uv\in\mathbb{S}^2$.
In addition to the supercooling temperature $T_c$ below which the isotropic state loses its stability, there are two other  characteristic temperatures for $w_{LdG}$: 
the nematic-isotropic transition temperature 
$\displaystyle\left(1+\frac{B^2}{27A_0C}\right)T_c
$
at which the nematic and the isotropic phase have the same energy,
and the superheating temperature
$
\displaystyle\left(1+\frac{B^2}{24A_0C}\right)T_c
$
above which the isotropic phase is the unique stationary point of $w_{LdG}$.
The resulting seven characteristic temperature regimes for $w_{LdG}$ are  discussed in detail by \cite{Turzi}.

\subsection{Degenerate states}
The procedure to derive  the two-dimensional free energy for nematic shells is performed in two subsequent steps: (a) we have to specialize the free energy to describe  degenerate planar distributions, where the eigenvector  of $\Mv$ with null eigenvalue coincides with $\bnu$;  then, (b) as for the OZF free energy, we approximate the three-dimensional free energy under the assumption of smallness of the parameter $\varepsilon$.  

To describe a degenerate anchoring throughout the shell,  let us suppose the nematic molecules are orthogonal to $\bnu$ and $\mv(p(u,v,\xi))= \mv(p_S(u,v))$. Since at each point the probability to find $\mv$ in the direction $\bnu$ is zero, it follows that $\Mv \bnu = {\bf 0}$.  This means that no isotropic spatial states are allowed.  Let us introduce $\nv$ and $\tav$ the eigenvectors of $\Mv$ orthogonal to $\bnu$.  We write the variance tensor in the form \citep{kralj:2011}
\ba
{\Mvs} =  \frac{1}{2}(\Iv-\bnu\ot\bnu) +  \lambda\left(\nv \ot \nv  -  \tav \ot \tav \right), \nonumber
\ea
where $\lambda \in[-\frac{1}{2},\frac{1}{2}]$. We recognize that two kinds of uniaxial states are allowed: (a) $\lambda =0$ then $\bnu$ is the optical axis and the molecules are randomly distributed orthogonally to $\bnu$; (b) $\lambda = \pm \frac{1}{2}$ and the molecules are perfectly ordered along a direction orthogonal to $\bnu$. The latter case coincides with the directory theory analyzed in the previous section.  Note that the sign of $\lambda$ is inessential since, the order  tensors associated with negative values of $\lambda$ and director $\nv$ coincide with the order tensors associated with the positive degree of order $-\lambda$	and	director  $\tav$.

An alternative and equivalent parameterization of the variance tensor is the following \citep{biscari:2006}:
\ba \label{bt}
{\Mvs}=  q \left(\nv \ot \nv  \right) + \frac{1}{2}(1-q)(\Iv-\bnu\ot\bnu),
\ea 
with $q = 2 \lambda \in \left[-1,1\right]$. 

It is worth noting  that this parameterization can be also obtained from the three-dimensional order parameter $\Mv$ by imposing $s=-\frac{1}{2}$,  by taking $\uv$ along the normal surface  and by choosing $\nv$ along one of the two tangential eigenvectors of $\Mv$. 

\subsection{Elastic energy}
Now, let us introduce the traceless tensor $\Qvs$, associated with $\Mvs$, in the usual way: 
$\Qvs = \Mvs - \frac{1}{3}\Iv$. This tensor differs from  the one of equation (6) in  \cite{kralj:2011}, which is indeed  the traceless tensor obtained by subtracting from $\Mvs$ one-half of the projector on $\esse$ (which is the identity on the tangent plane).

With the aim of adapting the elastic free energy to the case of degenerate states, we replace $\Qv$ by $\Qvs$. Since $\Qvs$ and $\Mvs$  (as well as $\Qv$ and $\Mv$) differ up to a constant, we have $\grad{\Qvs}=\grad\Mvs$; thus, in the elastic energy, $\Qvs$ can be replaced by $\Mvs$. 

 Furthermore,  by using the parameterization (\ref{bt}) and with the aid of equations (\ref{guno}-\ref{gtre}), the following identities hold:
\begin{align}\label{i1}
 \Ms_{ij,k} \Ms_{ij,k}&
\nonumber
=2q^2\left\{(\dv\nv)^2+|\nv\times\rot\nv|^2+(\rot\nv\cdot\nv)^2\right\} +\frac{1}{2}|\grad q|^2 \\
\nonumber
&+2\iota^{-2}(1-q)(\H-\kg \xi)[(1-q)\H+2q\cn-(1+q)\kg\ xi]\\
&-\iota^{-2}(1-q^2){\K},
\end{align}
\begin{align}\label{i2}
\Ms_{ij,k}\Ms_{ik,j}&=(\Ms_{ij,k}\Ms_{jk}-\Ms_{ij}\Ms_{jk,k})_{,i}+ \Ms_{ij,j}\Ms_{ik,k},
\end{align}
\begin{align} 
(\Ms_{ij,k}\Ms_{jk}-\Ms_{ij}\Ms_{jk,k})_{,i}&=
  2q\grad q\cdot[(\grad\nv)\nv-(\dv\nv)\nv]\\
  &+\dv\left\{\frac{1-q}{2\iota}[(1-q)\H+2q\cn-(1+q)\kg\ xi]\bnu\right\},
  \nonumber
\end{align}
\begin{align}\label{i3}
\Ms_{ij,j}\Ms_{ik,k}& \nonumber
=q^2\left[(\dv\nv)^2+|\nv\times\rot\nv|^2\right]+\frac{1}{4}|\grad q|^2 -q\nabla q\cdot [(\grad\nv)\nv-(\dv\nv)\nv]\\
&+\iota^{-2}(1-q)(\H-\kg \xi)[(1-q)\H+2q\cn-(1+q)\kg\ xi].
\end{align}
As for the director theory, in order  to obtain the elastic surface free energy, we expand  the volume free energy as a power series in the small parameter $\varepsilon$ and consider only the leading order term.  Thus,  by means  of Proposition \ref{elastic theorem} in Appendix B, we obtain
\begin{align}
\nonumber
W^{S}_{el} &=\int_\esse l_1\left[q^2\left(\kk^2+\kg^2\right)+\frac{1}{4}|\grads q|^2+\left(\H+q\frac{\cn-\ct}{2}\right)^2\right]\dd A \nonumber\\
&+\int_\esse l_2 q\grads q\cdot\left(\kg \tavs- \kk \nvs\right)\dd A-\int_\esse l_3(1-q^2)\kg \dd A
\nonumber\\
&-\int_\esse(l_1+l_2-4l_3)q^2\tg^2\dd A,
\label{el1}
\end{align}
 where $ l_1 = h(2L_1 + L_2)$,  $l_2 = h(L_2 + 2L_{24})$,  $l_3 = h(2L_1 + L_2 + L_{24})/2$.  In the next section we show that these elastic constants are subject to restrictions in order to guarantee the positiveness of the elastic free energy.

 In order to interpret the contributions of the different terms, we first examine the special case  where the perfect uniaxial nematic order ($q=1$ everywhere) is enforced on the entire shell.    
 Equation (\ref{el1}) reduces to
 $$
 W^{S}_{el}(q=1) = \int_\esse \left[l_1\left(\kk^2+\kg^2 + \cn^2\right)  - (l_1+l_2-4l_3)\tg^2\right]\dd A,
 $$
that represents a  Frank-like surface free energy (to be compared with equation (\ref{OZFshell})).  The  ratio between the twist and the splay constants can be tuned acting on the constant $l_i$. In particular, when $L_2=0$, then $4l_3 = 2l_1 + l_2$, and we retrieve the one constant approximation of the Frank's energy   (\ref{OZFshellone}). 
 
By denoting $\MS=q(\nvs\ot\nvs)+\frac{1}{2}(1-q)\Pv$  the restriction of $\Mvs$ to $\esse$,  the following identity holds
\ba
l_1\left(\H+q\frac{\cn-\ct}{2}\right)^2 = l_1 (\MS \cdot \Lv)^2;
\label{quadru}
\ea
the right  hand side of this identity is the quadrupolar coupling between the curvature tensor and the surface order tensor derived in  \cite{biscari:2006} employing  quasi-microscopic arguments.   When $q$ is different from zero, this term express the tendency of $\nvs$ to align along one of the two principal directions depending on the sign of the mean curvature.   
  
The energy term proportional to the square of surface gradient of $q$ clearly expresses the tendency of the nematic to arrange in states with constant order parameter. It is worth to note that, for topological reasons, states with non zero uniform $q$ are not always allowed.  This is the case of closed surfaces with the topology of the sphere.
 
The  term proportional to Gaussian curvature $K$ was already obtained in  \cite{kralj:2011}. It is a constant term only when $q$ is homogeneous on a fixed surface, by virtue of Gauss-Bonnet theorem.
 
Concerning the second integral of the right hand side of  (\ref{el1}), we find the following identity (see Appendix C) 
\ba
\int_\esse q (\grad_s q)\cdot (\kg \tavs - \kk \nvs) \dd A = \frac{1}{2}\int_{\pt S}  q^2 (\grads \alpha - \bom)\cdot  \dd \mathbf{l} + \frac{1}{2} \int_\esse q^2 K \dd A, 
\label{ide}
\ea
with $\alpha$ and $\bom$ as in previous section.
Thus, for closed shells,  the density free energy density associated with this term is even proportional to the Gaussian curvature.

\subsubsection{Restrictions on  the elastic coefficients}
The  positiveness of the free energy imposes suitable restrictions to the free energy coefficients.
Following the approach pursued in \cite{kralj:2011}, let us  decompose the surface elastic free energy density $w_{el}^\esse$ as follows 
\begin{equation}\label{els}
 w_{el}^\esse=w_{el1}^\esse +w_{el2}^\esse+w_{el3}^\esse,  
 \end{equation}
  with
 \begin{align}
w_{el1}^\esse =l_1\left[q^2(\kk^2 + \kg^2)+\frac{1}{4}|\grads q|^2\right]+l_2q\left(\kg\ tavs-\kk\nvs\right)\cdot\grads q,
\end{align}
\ba
w_{el2}^\esse = \frac{l_1}{4}\left[(1+q)^2\cn^2+2(1-q^2)\cn \ct+(1-q)^2\ct\right]-l_3(1-q^2)\cn \ct,
\ea
\ba\label{els3}
w_{el3}^\esse=\left[l_3-(l_1+l_2-3l_3)q^2\right]\tg^2,
\ea
where   the identity $\kg= \cn \ct-\tg^2$ has been used. 
 Then, we recognize that $w_{el1}^\esse = \vv_1 \cdot \Av_1 \vv_1$ and $w_{el2}^\esse = \vv_2 \cdot \Av_2 \vv_2$, with
$$
\mathbf{A}_1=\left(\begin{array}{cccc}
l_1 & l_2/2 & 0 & 0 \\
l_2/2 & l_1/4 & 0 & 0 \\
0 & 0 & l_1 & -l_2/2 \\
0 & 0 & -l_2/2 & l_1/4
\end{array}\right)
, \qquad
\mathbf{A}_2=\frac{1}{4}\left(\begin{array}{cc}
l_1 & l_1-2l_3 \\
l_1-2l_3 & l_1
\end{array}\right),
$$
$$\mathbf{v}_1:=(q\kg,\grads q\cdot\tavs,q\kk,\grads q\cdot\nvs), \qquad \mathbf{v}_2:=[(1+q)\cn,(1-q)\ct].$$
Hence, it can be  easily proved that $w_{el}^\esse\geq0$  if and only if 
$$
l_1\geq0,\quad | l_2 |\leq l_1, \quad 0\leq l_3\leq l_1, \quad l_1+ l_2\leq 4l_3,
$$
or, equivalently,
\ba
\label{eriLdG}
L_1\geq0, \quad 2L_1+L_2\geq0, \quad |L_{24}|\leq2L_1+L_2, \quad |L_2+2L_{24}|\leq2L_1+L_{2}.
\ea
By assuming $L_1>0$ and introducing the ratios  $\lambda_1=L_2/L_1$ and $\lambda_2=L_{24}/L_1$,  the admissible region in the  $(\lambda_1,\lambda_2)$-plane in which the surface elastic energy density  (\ref{el1}) is positive semidefinite, is sketched in  figure \ref{fig:dominio}. It is worth noting that the domain in which the elastic free energy density  (\ref{els}) is positive semidefinite contains the domain of nonnegativeness of the surface energy density introduced in \cite{kralj:2011}.  This in turn contains the domain of nonnegativeness of the elastic energy  density  (\ref{elastic energy}).

\begin{figure}
\centerline{\includegraphics[width=8cm]{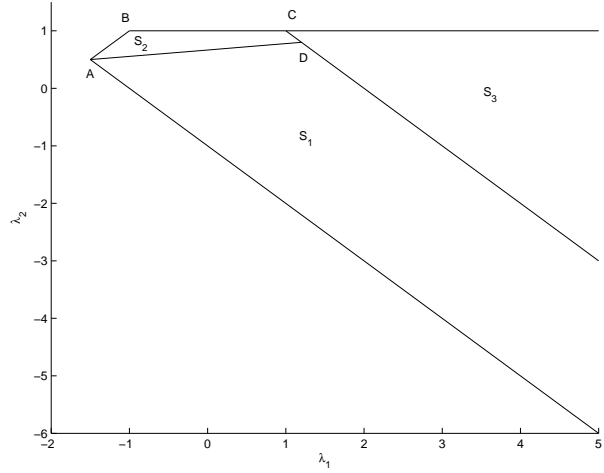}}
\label{fig:dominio}
\caption{\label{fig:dominio} We have set $\lambda_1=L_2/L_1$, $\lambda_2=L_{24}/L_1$.  $S_1$ is the region in which  the elastic energy  density in (\ref{elastic energy}) is nonnegative (see \cite{longa:1987}); $S_1\cup S_2$ represent the domain in which the surface elastic free energy in \cite{kralj:2011} is nonnegative; finally $S_1\cup S_2 \cup S_3$ is the region where inequalities the (\ref{eriLdG}) hold. $A\equiv(-3/2,1/2)$, $B\equiv(-1,1)$, $C\equiv(1,1)$, $D\equiv(6/5,4/5)$.}
\end{figure}

\subsection{Landau-de Gennes potential}
Let us consider the Landau-de Gennes free energy density (\ref{enldg}),  where $\Qv = \Qvs$. A straightforward calculation gives
$$
\tr(\Qvs^2)=\frac{1}{6}+\frac{1}{2}q^2,\quad \tr(\Qvs^3)=-\frac{1}{36}+\frac{1}{4}q^2.
$$
Following the same arguments given in  Appendix B,  we readily derive the surface Landau-de Gennes free energy   
$$
W_{LdG}\approx W_{LdG}^\esse=\int_\esse\left(d+\frac{a}{4}q^2+\frac{c}{8}q^4\right)\dd A \quad \mathrm{for} \:\varepsilon\ll1,
$$
where
$$
d= h\left[F(A,B,C)+\frac{A}{12}+\frac{B}{108}+\frac{C}{144}\right], \quad a = a_0\frac{T-T_c^*}{T_c},\quad a_0=hA_0,$$ 
$$
c=h\frac{C}{2},\quad T_c^*=\left(1+\frac{B}{3A_0}-\frac{C}{6A_0}\right)T_c.
$$
We then obtain  a Landau-de Gennes potential with two constants in which the  cubic term  vanishes.  An analogous expression is proposed in \cite{biscari:2006, kralj:2011}. It is worth noting that,  homogenous states with $q\neq 0$ are allowed only on surfaces with zero Euler-Poincar\'e chracteristic.  In fact, only in this case it is possible to define a critical temperature that  generally depends on the shell curvature. 
\section{Concluding remarks}
We have deduced  the two-dimensional versions of Frank and Landau-de Gennes free energies needed   to treat the equilibrium of thin nematic films, coating curved surfaces. These models have been obtained as limiting cases of the respective three-dimensional models.  The formalism proposed applies to rigid shells as well as to flexible surfaces with two-dimensional nematic order. Obviously, in the latter case additional energy terms are required to describe the elasticity of the shell. The problem of equilibrium can be framed in the general variational scheme proposed  in \cite{napoli:2010}. However, the resulting equations for this complex problem are strongly non linear and demand a numerical treatment. 

Our rigorous procedure predicts the existence of new terms in the free energy, with respect to earlier models. The physical interpretation of these extra terms is widely discussed in Sect. 3 and Sect. 4. The key results of our analysis are as follows: 
 \begin{itemize}
 \item [(i)] In the context of the director theory for curved nematic thin films, the twist free energy does not vanish. This free energy, coupled with the term proportional to $\cn^2$, expresses the tendency of the molecules to align along the principal direction of the surface with minimum curvature. Thus, the extrinsic geometry of the shell influences the molecular alignment  in  agreement  with the results announced in  \cite{santangelo:2011}. In a forthcoming work, we show how the twist  term influences the stability of a nematic on a toroidal surface.   
\item [(ii)]  In the context of  Landau-de Gennes theory, we establish a coherent framework to develop  a two-dimensional order-tensor theory. As a result, we obtain the coupling term (\ref{quadru}). This  term has been already proposed in  \cite{biscari:2006}, but it required  an additional phenomenological constant in the model.  By contrast, since we deduce that  the coefficient  of this  energy is  the Frank's constant $k_1$,  no further  phenomenological constants should be introduced. We notice that, within the model proposed \cite{kralj:2011}, this term does not appear; this implies the counterintuitive fact that the biaxiality axes can be interchanged without affecting the energy. 
\end{itemize}
Our approach offers the two-fold advantage of being based on well-established theories and, at same time, to avoid the proliferation of phenomenological coefficients in the free energy expression.  Therefore, our models describe in an {\it economical} and {\it exhaustive} manner the equilibrium configuration of in-plane curved nematics. Obviously, our procedure can be extended to more complicated models as that proposed in \cite{longa:1987}.

We believe that the results outlined in this paper are the basis to  study the arrangement of two-dimensional curved nematics. We envisage a series of future  studies to establish the influence of external actions (temperature, electric or magnetic fields), of the shell geometry, and  of the material coefficients on the  nematic shell texture.

\section*{Acknolodgements}
{The authors would like to thank Stefano Turzi for useful discussions on the topics of this paper.}

\appendix

\section{ Derivation of the spatial gradients of $\bnu$, $\ev_1$, $\ev_2$ and $\nv$} 
Let $\phiv=\phiv(u,v)$ an orthogonal parameterization  of $\esse$ such that $$\frac{\phiv_{,u}}{|\phiv_{,u}|}=\ev_{1s} \quad \textrm{and} \quad \frac{\phiv_{,v}}{|\phiv_{,v}|}=\ev_{2s}.$$
 Then, for any fixed $\xi\in[-h/2,h/2]$,
$
\phiv_\xi=\phiv(u,v)+\xi\bnus(u,v)
$
is an orthogonal parameterization of $\esse_\xi$ such that
$$
\phiv_{\xi,u}=(1-\xi c_{1s})\phiv_{,u} \quad \textrm{and} \quad \phiv_{\xi,v}=(1-\xi c_{2s})\phiv_{,v}.
$$
As  consequences of assumption (\ref{hl}) , $\{\ev_1,\ev_2\}$, with
\begin{equation}\label{e1}
\ev_1(p):=\frac{\phiv_{\xi,u}}{|\phiv_{\xi,u}|}=\frac{\phiv_{,u}}{|\phiv_{,u}|}=\ev_{1s}(p_\esse) \quad \forall p\in V
\end{equation}
and
\begin{equation}\label{e2}
\ev_2(p):=\frac{\phiv_{\xi,v}}{|\phiv_{\xi,v}|}=\frac{\phiv_{,v}}{|\phiv_{,v}|}=\ev_{2s}(p_\esse) \quad \forall p\in V,
\end{equation}
is a local orthonormal basis of the space of tangent vectors $\mathfrak{X}(\esse_\xi)$,
whereas
\begin{equation}\label{normale}
\bnu(p):=\frac{\phiv_{\xi,u}\times\phiv_{\xi,v} }{|\phiv_{\xi,u}\times\phiv_{\xi,v}|}=\frac{\phiv_{,u}\times\phiv_{,v} }{|\phiv_{,u}\times\phiv_{,v}|}=\bnus(p_\esse)  \quad \forall p\in V
\end{equation}
is  the unit normal vector field on $\esse_\xi$.
We now introduce the following quantities
\begin{align*}
&e_\xi=-{\bnus}_{,u}\cdot\phiv_{\xi,u}=c_{1s}(1-\xi c_{1s})\phiv_{,u}\cdot\phiv_{,u}=c_{1s}(1-\xi c_{1s})E,\\
\vspace{2mm}
&f_\xi=-{\bnus}_{,u}\cdot\phiv_{\xi,v}=0=-{\bnus}_{,v}\cdot\phiv_{\xi,u},\\
\vspace{2mm}
&g_\xi=-{\bnus}_{,v}\cdot\phiv_{\xi,v}=c_{2s}(1-\xi c_{2s})\phiv_{,v}\cdot\phiv_{,v}=c_{2s}(1-\xi c_{2s})G,\\
\vspace{2mm}
&E_\xi=\phiv_{\xi,u}\cdot\phiv_{\xi,u}=(1-\xi c_{1s})^2\phiv_{,u}\cdot\phiv_{,u}=(1-\xi c_{1s})^2E,\\
\vspace{2mm}
&F_\xi=\phiv_{\xi,u}\cdot\phiv_{\xi,v}=(1-\xi c_{1s})(1-\xi c_{2s})\phiv_{,u}\cdot\phiv_{,v}=0,\\
\vspace{2mm}
&G_\xi=\phiv_{\xi,v}\cdot\phiv_{\xi,v}=(1-\xi c_{2s})^2\phiv_{,v}\cdot\phiv_{,v}=(1-\xi c_{2s})^2G,
\end{align*}
where $E=\phiv_{,u}\cdot\phiv_{,u}$ and $G=\phiv_{,v}\cdot\phiv_{,v}$.

We first derive the gradient of $\bnu$. From (\ref{normale})  it follows that
\begin{equation}\label{passo1}
(\nabla\bnu)\bnu=\mathbf{0}, 
\end{equation}
 and, since it is a unit vector field, 
 \begin{equation}
 \bnu \cdot (\nabla\bnu)\ev_i = 0 \quad \forall i=1,2. \nonumber
 \end{equation}
Moreover, for any fixed $\xi$, $-\grad\bnu$  restricted to the space of tangent vectors $\mathfrak{X}(\esse_\xi)$ represents the extrinsic curvature tensor  of $\esse_\xi$. Therefore, following  \citet{docarmo}:
 \begin{equation}\label{passo3}
 \left.\begin{array}{ll}
 \ev_1 \cdot (\grad \bnu)\ev_1=-\displaystyle\frac{f_\xi F_\xi-e_\xi G_\xi}{E_\xi G_\xi-F_\xi^2}=-\frac{c_{1s}}{1-\xi c_{1s}},\\
 [5mm]
 \ev_1 \cdot (\grad \bnu)\ev_2=-\displaystyle\frac{g_\xi F_\xi-f_\xi G_\xi}{E_\xi G_\xi-F_\xi^2}=0,\\
 [5mm]
 \ev_2 \cdot (\grad \bnu)\ev_1=-\displaystyle\frac{e_\xi F_\xi-f_\xi E_\xi}{E_\xi G_\xi-F_\xi^2}=0,\\
 [5mm]
\ev_2 \cdot  (\grad \bnu)\ev_2=-\displaystyle\frac{f_\xi F_\xi-g_\xi E_\xi}{E_\xi G_\xi-F_\xi^2}=-\frac{c_{2s}}{1-\xi c_{2s}},
\end{array}\right.
\end{equation}
by which we deduce that
$
\ev_1
$ and $\ev_2$ are the tangent principal directions on $\esse_\xi$.
Finally, (\ref{passo1})-(\ref{passo3}) yield (\ref{guno}).

Let us now calculate $\grad\ev_i$ $(i=1,2)$. From (\ref{e1}), (\ref{e2}) and since $\ev_i$ ($i=1,2$) are unit vector fields, we deduce that
\begin{equation}\label{1e}
(\nabla\ev_i)\bnu=\mathbf{0}=(\nabla \ev_i)^T\ev_i \quad \forall i=1,2.
\end{equation}
Next, since $\{\ev_1,\ev_2,\bnu\}$ is a local orthonormal basis
\begin{equation}\label{2e}
(\nabla\ev_i)^T\ev_j=-(\nabla\ev_j)^T\ev_i \quad \forall i,j=1,2,\:\: i\neq j,
\end{equation}
and
\begin{equation}\label{3e}
\bnu \cdot (\nabla \ev_i)\ev_j = -\ev_j \cdot (\nabla\bnu)^T\ev_i =\delta_{ij}\frac{c_i}{1-\xi c_i} \quad \forall i,j=1,2,
\end{equation}
where $\delta_{ij}$ denotes the Kronecker symbol. By means of  (\ref{2e}),
\begin{equation}\label{een}
\left.\begin{array}{ll}
\ev_2 \cdot (\nabla\ev_1)\ev_1  =  - \ev_1 \cdot (\nabla\ev_2)\ev_1  = \kappa_1(\xi),\\
[3mm]
\ev_2 \cdot (\nabla\ev_1)\ev_2       =-\ev_1 \cdot (\nabla\ev_2)\ev_2 = \kappa_2(\xi),
\end{array}\right.
\end{equation}
where $ \kappa_1(\xi)$ and $ \kappa_2(\xi)$
 are the geodesic curvatures of the lines of curvature on $\esse_\xi$. Hence, by following \cite{docarmo} and since the surface gradient of a scalar-valued function $f$ defined in a neighborhood of $\esse$ may be written as
\begin{equation}\label{grads}
\grads f= \frac{f_{,u}}{\sqrt{E}}\ev_{1 s}+\frac{f_{,v}}{\sqrt{G}}\ev_{2 s},
\end{equation}
the geodesic curvatures of the lines of curvature on $\esse_\xi$ are found to be
\begin{align}\label{chr1}
 \kappa_1(\xi)=-\frac{E_{\xi,v}}{2E_\xi\sqrt{G_\xi}}&=-\frac{E_{,v}}{2(1-\xi c_{2s})E\sqrt{G}}+\frac{\xi c_{1,v}}{(1-\xi c_{1s})(1-\xi c_{2s})\sqrt{G}}\\
\nonumber
&=\frac{\kappa_{1s}}{1-\xi c_{2s}}+\frac{\xi \grads c_{1s}\cdot \ev_{2 s}}{(1-\xi c_{1s})(1-\xi c_{2s})},
\end{align}
and 
\begin{align}\label{chr2}
 \kappa_2(\xi)=\frac{G_{\xi,u}}{2G_\xi\sqrt{E_\xi}}=&\frac{G_{,u}}{2(1-\xi c_{1s})G\sqrt{E}}
-\frac{\xi c_{2,u}}{(1-\xi c_{1s})(1-\xi c_{2s})\sqrt{E}}\\
\nonumber
&=\frac{\kappa_{2s}}{1-\xi c_{1s}}-\frac{\xi \grads c_{2s}\cdot \ev_{1 s}}{(1-\xi c_{1s})(1-\xi c_{2s})},
\end{align}
where 
$$
\kappa_{1s}=-\frac{E_{,v}}{2E\sqrt{G}} \quad \mathrm{and}\quad \kappa_{2s}=\frac{G_{,u}}{2G\sqrt{E}}
$$
 are the geodesic curvatures of the lines of curvature on $\esse$.
Therefore,  (\ref{1e})--(\ref{chr2}) give (\ref{gdue}) and (\ref{gtre}).

We are now in position to derive the gradient of the director field $\nv$. Since $\nv$ is a unit vector field that does not vary with $\xi$ and is pointwise orthogonal to $\bnu$, we get 
\begin{equation}\label{n0}
(\nabla\nv)\bnu=\mathbf{0}=(\nabla\nv)^T\nv.
\end{equation}
 Next, we  introduce the angle $\alpha$ that $\nv$ form with $\ev_1$  so that we may write 
\begin{equation}\label{n1}
\nv=\cos \alpha \ev_1+\sin\alpha \ev_2, \quad \tav=\bnu\times\nv=-\sin\alpha \ev_1+\cos\alpha \ev_2
\end{equation}
and  
\begin{equation}\label{n2}
\grad\nv=-\sin\alpha\ev_1\ot\grad\alpha+\cos\alpha\nabla\ev_1+\cos\alpha\ev_2\ot\grad\alpha+\sin\alpha\grad\ev_2.
\end{equation}
 Since $\nv$ and $\ev_1$ are constant throughout the thickness, also the scalar field $\alpha$ satisfies the equality  $\alpha(p)=\alpha(p_\esse)$ for all $p\in V$. Therefore, in view of (\ref{grads})  the spatial gradient of the scalar field $\alpha$ is 
 \begin{align}
 \nonumber
 \grad\alpha&=\frac{\alpha_{,u}}{(1-\xi c_{1s})\sqrt{E}}\ev_1+\frac{\alpha_{,v}}{(1-\xi c_{2s})\sqrt{G}}\ev_2\\
 \nonumber
 &=\frac{\grads\alpha\cdot\ev_{1 s}}{1-\xi c_{1s}}\ev_1+\frac{\grads\alpha\cdot\ev_{2 s}}{1-\xi c_{2s}}\ev_2
 \end{align}
Thus,
\begin{align}\label{nun}
\nonumber
\bnu \cdot (\grad\nv)\nv &=-\nv \cdot (\grad\bnu)\nv=\frac{c_{1s}\cos^2\alpha+c_{2s}\sin^2\alpha-\xi c_{1s}c_{2s}}{(1-\xi c_{1s})(1-\xi c_{2s})}\\
&=\frac{\cn-\xi K}{1-2\xi\H+\xi^2\K},
\end{align}
\begin{equation}\label{nunt}
\bnu \cdot (\grad\nv)\tav=-\nv \cdot (\grad \bnu)\tav=\frac{(c_{2s}-c_{1s})\sin\alpha\cos\alpha}{(1-\xi c_{1s})(1-\xi c_{2s})}=-\frac{\tg}{1-2\xi\H+\xi^2\K},
\end{equation}
\begin{align}\label{tn}
\nonumber
\tav \cdot (\grad\nv)\nv&=\frac{(\grads\alpha\cdot\ev_{1 s})\cos\alpha+(\grads\alpha\cdot\ev_{2 s})\sin\alpha+\kappa_{1s}\cos\alpha+\kappa_{2s}\sin\alpha}{(1-\xi c_{1s})(1-\xi c_{2s})}\\
\nonumber
&-\xi\frac{c_{1s}\kappa_{1s}\cos\alpha+c_{2s}\kappa_{2s}\sin\alpha+\grads c_{2s}\cdot\ev_{1 s}\sin\alpha-\grads c_{1s}\cdot\ev_{2 s}\cos\alpha}{{(1-\xi c_{1s})(1-\xi c_{2s})}}\\
\nonumber
&-\xi\frac{c_{2s}\cos\alpha\grads\alpha\cdot\ev_{1 s}+c_{1s}\sin\alpha\grads\alpha\cdot\ev_{2 s}}{(1-\xi c_{1s})(1-\xi c_{2s})}\\
\nonumber
& = \frac{\grads\alpha\cdot\nvs+\kappa_{1s}\cos\alpha+\kappa_{2s}\sin\alpha-\xi\dvs(c_{2s}\sin\alpha\ev_{1 s}-c_{1s}\cos\alpha\ev_{2 s})}{1-2\H\xi+\kg\ xi^2}\\
&= \frac{\kg-\xi\bnus\cdot\rots(\Lv\nvs)}{1-2\H\xi+\kg\ xi^2}
\end{align}
and
\begin{align}\label{tt}
\nonumber
\tav \cdot (\grad\nv)\tav&=\frac{-(\grads\alpha\cdot\ev_{1 s})\sin\alpha+(\grads\alpha\cdot\ev_{2 s})\cos\alpha-\kappa_{1s}\sin\alpha+\kappa_{2s}\cos\alpha}{(1-\xi c_{1s})(1-\xi c_{2s})}\\
\nonumber
&-\xi\frac{-c_{1s}\kappa_{1s}\sin\alpha+c_{2s}\kappa_{2s}\cos\alpha+\grads c_{2s}\cdot\ev_{1 s}\cos\alpha+\grads c_{1s}\cdot\ev_{2 s}\sin\alpha}{{(1-\xi c_{1s})(1-\xi c_{2s})}}\\
\nonumber
&+\xi\frac{c_{2s}\sin\alpha\grads\alpha\cdot\ev_{1 s}-c_{1s}\cos\alpha\grads\alpha\cdot\ev_{2 s}}{(1-\xi c_{1s})(1-\xi c_{2s})}\\
\nonumber
& = \frac{\grads\alpha\cdot\tavs-\kappa_{1s}\sin\alpha+\kappa_{2s}\cos\alpha-\xi\dvs(c_{2s}\cos\alpha\ev_{1 s}+c_{1s}\sin\alpha\ev_{2 s})}{1-2\H\xi+\kg\ xi^2}\\
&= \frac{\kk -\xi\bnus\cdot\rots(\Lv\tavs)}{1-2\H\xi+\kg\ xi^2},
\end{align}
where $\nvs=\cos\alpha\ev_{1 s}+\sin\alpha\ev_{2 s}$ and $\tavs=\bnus\times\nvs$ are the restrictions of $\nv$ and $\tav$ on $\esse$, respectively, and $\Lv=c_{1s}\ev_{1 s}\ot\ev_{1 s}+c_{2s}\ev_{2 s}\ot\ev_{2 s}$ is the extrinsic curvature tensor on $\esse$. The quantities 
$$
\cn=c_{1s}\cos^2\alpha+c_{2s}\sin^2\alpha \quad \mathrm{and} \quad c_{\tavs}=c_{1s}\sin^2\alpha +c_{2s}\cos^2\alpha
$$
are the normal curvatures of the flux lines of $\nvs$ and $\tavs$, respectively, whereas 
$$
\tg=(c_{1s}-c_{2s})\sin\alpha\cos\alpha
$$ 
is the geodesic torsion of the flux lines of $\nvs$. 
In deriving (\ref{tn}) and (\ref{tt}) we have made use of  the Lioville's formula (see \cite{docarmo} page 253) for the calculation of the geodesic curvatures $\kg$ and $\kk$, \textit{i.e.}
$$
\kg= \grads\alpha\cdot\nvs+\kappa_{1s}\cos\alpha+\kappa_{2s}\sin\alpha, \quad \kk=\grads\alpha\cdot\tavs-\kappa_{1s}\sin\alpha+\kappa_{2s}\cos\alpha,
$$
and have employed the identity
\begin{equation}\label{ida1}
\dvs(\bnus\times\mathbf{u})=-\bnus\cdot\rots\mathbf{u}
\end{equation}
that holds true for any smooth field $\uv$ defined on $\esse$.
We may then conclude that (\ref{n0})--(\ref{tt}) yield (\ref{gradn}).

\section{ Derivation of $W_{OZF}^S$ and $W_{el}^S$}
In this section we shall derive the approximations of the energies (\ref{OZF}) and (\ref{elastic energy}) that are valid for a homogeneous nematic whenever $  \varepsilon\ll1$.

\begin{proposition}\label{elastic theorem}
Let $\nv$ and $q$  be  smooth fields defined on  $V$. Assume $\nv$ to be a unit vector field such that
$$
\nv(p)\cdot\bnu(p)=0 \quad \forall p\in V
$$
and
\begin{equation}\label{ipotesi n}
\nv[p_\esse+\xi\bnus(p_\esse)]=\nv(p_\esse) \quad \forall p_\esse\in\esse,\forall \xi\in[-h/2,h/2],
\end{equation}
and $q$ a scalar-valued field such that
\begin{equation}\label{ipotesi q}
q[p_\esse+\xi\bnus(p_\esse)]=q(p_\esse)\in[-1,1] \quad \forall p_\esse\in\esse,\forall \xi\in[-h/2,h/2].
\end{equation}
Then, denoting by $\mathrm{vol}(V)$  the volume of $V$,
\begin{align}\label{lim1}
\nonumber
\lim_{  \varepsilon \rightarrow0}&\int_V\frac{\Ms_{ij,k}\Ms_{ij,k}}{\mathrm{vol}(V)}\dd V=\int_\esse \frac{q^2\left[(\dvs\nvs)^2+|\nvs\times\rots\nvs|^2+(\rots\nvs\cdot\nvs)^2\right]}{\mathrm{area}(\esse)}\dd A\\
&+\int_{\esse}\frac{\displaystyle\frac{|\grads q|^2}{2}+2(1-q)\H[(1-q)\H+2q \cn]-(1-q^2)\K}{\mathrm{area}(\esse)}\dd A,
\end{align}
\begin{align}\label{lim2}
&\lim_{  \varepsilon \rightarrow0}\int_V\frac{\Ms_{ij,j}\Ms_{ik,k}}{\mathrm{vol}(V)}\dd V=\int_\esse\frac{ q^2\left[(\dvs\nvs)^2+|\nvs\times\rots\nvs|^2\right]}{\mathrm{area}(\esse)}\dd A\\
\nonumber
&+\int_{\esse}\frac{\displaystyle\frac{|\grads q|^2}{4}-q\grads q \cdot[(\rots\nvs\cdot\bnus)\tav-(\dvs\nvs)\nvs]+(1-q)\H[(1-q)\H+2q \cn]}{\mathrm{area}(\esse)}\dd A,
\end{align}
\begin{align}\label{limit3}
\nonumber
\lim_{  \varepsilon \rightarrow0}\int_V\frac{(\Ms_{ij,k}\Ms_{jk}-\Ms_{ij}\Ms_{jk,k})_{,i}}{\mathrm{vol}(V)}\dd V&=\int_\esse\frac{ 2q\grads q \cdot[(\rots\nvs\cdot\bnus)\tavs-(\dvs\nvs)\nvs]}{\mathrm{area}(\esse)}\dd A\\
&-\int_\esse\frac{(1-q^2)}{2\mathrm{area}(\esse)}\kg\ d A,
\end{align}
where 
$$
\Ms_{ij}=qn_in_j+\frac{1}{2}(1-q)(\delta_{ij}-\nu_i\nu_j).
$$
\end{proposition}

We observe that
\ba \label{volume}
\mathrm{vol}(V)=h\left[\mathrm{area}(\esse)+\displaystyle\frac{h^2}{12}\int_\esse\kg\ d A\right].
\ea
With the aid of equation  (\ref{divergenza})
\begin{align}
\nonumber
\int_V \frac{q^2(\dv\nv)^2}{\mathrm{vol}(V)}\dd V&=\int_{-h/2}^{h/2}\dd\xi\int_{\esse_\xi}\frac{q^2 \left[\kk-\xi\bnus\cdot\rots(\Lv\tavs)\right]^2}{{\iota^2\mathrm{vol}(V)}}\dd A\\
\nonumber
&
=\int_{-h/2}^{h/2}\dd\xi\int_\esse \frac{ q^2\left[\kk-\xi\bnus\cdot\rots(\Lv\tavs)\right]^2}{{\iota\mathrm{vol}(V)}}\dd A\\
\nonumber
&=\int_\esse\left\{\int_{-h/2}^{h/2}\frac{q^2 \left[\kk-\xi\bnus\cdot\rots(\Lv\tavs)\right]^2}{{\iota\mathrm{vol}(V)}}\ dd \xi\right\} \dd A.
\end{align}
Since $q$, $\kk$ and $\bnus\cdot\rots(\Lv\tavs)$  do not depend on $\xi$,  by means of (\ref{volume}) we deduce that
$$
\int_{-h/2}^{h/2}\frac{q^2[\kk-\xi\bnus\cdot\rots(\Lv\tavs)]^2}{\iota h\left(\mathrm{area}(\esse)+\displaystyle\frac{h^2}{12}\int_\esse\kg\ d A\right)}\dd \xi\rightarrow \frac{q^2\kk^2}{\mathrm{area}(\esse)}
$$
uniformly in $\esse$ as $  \varepsilon\rightarrow0$.
Therefore, recalling $(\ref{dvsrots})_1$, 
\begin{equation}\label{part1}
\lim_{  \varepsilon\rightarrow0}\int_V \frac{q^2(\dv\nv)^2}{\mathrm{vol}(V)} \dd V=\int_\esse \frac{q^2(\dvs\nvs)^2}{{\mathrm{area}(\esse)}}\dd A.
\end{equation}

We now use equation (\ref{rotore}) to obtain 
\ba
\int_V\frac{q^2(\rot\nv\cdot\nv)^2}{\mathrm{vol}(V)}\dd V= \int_{\esse}\left[\int_{-h/2}^{h/2}\frac{q^2\tv^2}{\iota {\mathrm{vol}(V)}}\dd \xi\right]\dd A
\nonumber
\ea
and
\begin{align}
\nonumber
\int_V \frac{q^2|\nv\times\rot\nv|^2}{\mathrm{vol}(V)} \dd V=
\int_{\esse}\left\{\int_{-h/2}^{h/2}\frac{q^2(\cn-\kg\ xi)^2+[\kg-\xi\bnus\cdot\rots(\Lv\nvs)]^2}{\iota\mathrm{vol}(V)}\dd \xi\right\}\dd A.
\end{align}
Considering that $q$, $\cn$, $\tg$, $\kg$  and $\bnus\cdot\rots(\Lv\nvs)$ do not depend on $\xi$, by means of (\ref{volume}), we have 
\[
\int_{-h/2}^{h/2}\frac{q^2\tg^2 }{\iota h\left(\mathrm{area}(\esse)+\displaystyle\frac{h^2}{12}\int_\esse\kg\ d A\right)}\dd \xi\rightarrow \frac{q^2\tg^2}{\mathrm{area}(\esse)}\quad \textrm{uniformly in $\esse$ as $  \varepsilon\rightarrow0$}
\]
and
\[
\int_{-h/2}^{h/2}q^2\frac{(\cn-\kg\ xi)^2+[\kg-\xi\bnus\cdot\rots(\Lv\nvs)]^2}{\iota h\left(\mathrm{area}(\esse)+\displaystyle\frac{h^2}{12}\int_\esse\kg \dd A\right)} {\rm d} \xi\rightarrow \frac{q^2(\cn^2 + \kg^2)}{\mathrm{area}(\esse)}
\]
 uniformly in $\esse$ as $  \varepsilon\rightarrow0$.
Thus, from (\ref{dvsrots}) we deduce that
\begin{equation}\label{part2}
\lim_{  \varepsilon\rightarrow0}\int_V \frac{q^2(\nv\cdot\rot\nv)^2}{\mathrm{vol}(V)}\dd V=\int_\esse \frac{q^2(\nvs\cdot\rots\nvs)^2}{\mathrm{area}(\esse)}\dd A
\end{equation}
and
\begin{equation}\label{part3}
\lim_{  \varepsilon \rightarrow0}\int_V\frac{q^2|\nv\times\rot\nv|^2}{\mathrm{vol}(V)}\dd V=\int_\esse \frac{q^2|\nvs\times\rots\nvs|^2}{\mathrm{area}(\esse)}\dd A.
\end{equation}

By following the same arguments which lead to  (\ref{part1})--(\ref{part3}) and by taking into account that
$$
\grad q=\frac{\grads q\cdot\ev_{1s}}{1-\xi c_{1s}}\ev_1+\frac{\grads q\cdot\ev_{2s}}{1-\xi c_{2s}}\ev_2,
$$
 one can easily prove that
\begin{equation}\label{part4}
\lim_{  \varepsilon \rightarrow0}\int_V\frac{|\grad q|^2}{\mathrm{vol}(V)}\dd V=\int_\esse\frac{|\grads q|^2}{\mathrm{area}(\esse)}\dd A,
\end{equation}
\begin{align}\label{part5}
\nonumber
\lim_{  \varepsilon \rightarrow0}\int_V&\frac{(1-q)(\H-\kg \xi)}{\iota^2\mathrm{vol}(V)}[(1-q)\H+2q\cn-(1+q)\kg\ xi]\\
&=\int_\esse\frac{(1-q)\H}{\mathrm{area}(\esse)}[(1-q)\H+2q \cn]\dd A,
\end{align}
\begin{equation}\label{part6}
\lim_{  \varepsilon\rightarrow0}\int_V\frac{(1-q^2)\K}{\iota^2\mathrm{vol}(V)}\dd V=\int_\esse\frac{(1-q^2)\K}{\mathrm{area}(\esse)}\dd A,
\end{equation}
\begin{align}\label{part7}
\nonumber
\lim_{  \varepsilon\rightarrow0}\int_V&\frac{q\grad q\cdot[(\nabla\nv)\nv-(\dv\nv)\nv]}{\mathrm{vol}(V)}\dd V\\
&=\int_\esse\frac{q\grads q\cdot[(\rots\nvs\cdot\bnus)\tavs-(\dvs\nvs)\nvs]}{\mathrm{area}(\esse)}\dd A.
\end{align}

Now, let us assume now that $\esse$ is a regular surface whose boundary  $\partial\esse$ is a regular curve, and let $\tv$ be the tangent unit vector field to $\partial \esse$. Then, the normal unit vector field  to the surface 
$$
\esse_l:=\left\{p_{\partial\esse}+\xi\bnus(p_{\partial\esse}):p_{\partial\esse}\in\partial\esse,\xi\in[-h/2,h/2]\right\}
$$
 is
$$
\mathbf{N}=\frac{(\tv-\xi\Lv\tv)\times\bnu}{|(\tv-\xi\Lv\tv)\times\bnu|}.
$$
Therefore, by means of the divergence theorem we deduce that
\begin{align}
\nonumber
\int_V\dv&\Big\{\iota^{-1}(1-q)[(1-q)\H+2q\cn-(1+q)\kg\ xi]\bnu\Big\}\dd V\\
\nonumber
&=\int_{\esse_{h/2}}\frac{1-q}{1-\H h+\kg h^2/4}\left[(1-q)\H+2q\cn-(1+q)\kg \frac{h}{2}\right]\dd A\\
\nonumber
&-\int_{\esse_{-h/2}}\frac{1-q}{1+\H h+\kg h^2/4}\left[(1-q)\H+2q\cn+(1+q)\kg \frac{h}{2}\right]\dd A\\
\nonumber
&+\int_{\esse_l}\iota^{-1}(1-q)[(1-q)\H+2q\cn-(1+q)\kg\ xi]\bnu\cdot\mathbf{N}\dd A\\
\nonumber
&=\int_{\esse}(1-q)\left[(1-q)\H+2q\cn-(1+q)\kg \frac{h}{2}\right]\dd A\\
\nonumber
&-\int_{\esse}(1-q)\left[(1-q)\H+2q\cn+(1+q)\kg \frac{h}{2}\right]\dd A=-\int_\esse h(1-q^2)\kg\ d A.
\end{align}
By means of (\ref{volume}) we may conclude that
\begin{align}\label{part8}
\lim_{  \varepsilon \rightarrow0}&\int_V\dv\Big\{\frac{(1-q)}{\iota\mathrm{vol}(V)}[(1-q)\H+2q\cn-(1+q)\kg\ xi]\bnu\Big\}\dd V\\
\nonumber
&=-\lim_{  \varepsilon \rightarrow0}\int_\esse \frac{(1-q^2)\K}{\mathrm{area}(\esse)+\displaystyle\frac{h^2}{12}\int_\esse\kg\ d A}\dd A=-\int_\esse \frac{(1-q^2)\K}{\mathrm{area}(\esse)}\dd A.
\end{align}
We arrive at (\ref{part8}) also whenever $\esse$ is a geometrically closed surface, i.e., $\partial\esse=\emptyset$.

Finally, equations (\ref{lim1})--(\ref{limit3}) immediately follows from (\ref{i1})--(\ref{i3}) and (\ref{part1})--(\ref{part8}).

From (\ref{volume}) it follows that
\begin{equation}\label{limitvolume}
\lim_{  \varepsilon\rightarrow0}\mathrm{vol}(V)=h\mathrm{area}(\esse).
\end{equation}

As an immediate consequence of Proposition \ref{elastic theorem} and (\ref{limitvolume}), we have 
\begin{align}
\nonumber
W_{el}(\grad \Qv, \Qv)&=\mathrm{vol}(V)\Bigg\{\int_V\frac{L_1\Ms_{ij,k}\Ms_{ij,k}+L_2\Ms_{ij,j}\Ms_{ik,k}}{\mathrm{vol}(V)}\dd V\\
\nonumber
&+\int_V\frac{L_{24}(\Ms_{ij,k}\Ms_{jk}-\Ms_{ij}\Ms_{jk,k})_{,i}}{\mathrm{vol}(V)}\dd V\Bigg\}\approx W_{el}^S \quad \mathrm{for} \:\:  \varepsilon\ll 1,
\end{align}
with $W_{el}^\esse$ as in (\ref{el1}).

On taking $q\equiv1$ in (\ref{part1})--(\ref{part3}) we have the following
\begin{proposition}\label{Frank theorem}
Let $\nv$ be a smooth unit vector field defined on $V$ such that $$\nv(p)\cdot \bnu(p)=0 \quad \forall p\in V$$
and
$$
\nv(p_\esse+\xi\bnus)=\nv(p_\esse) \quad \forall p_\esse\in\esse,\forall \xi\in[-h/2,h/2].
$$
Then
\ba
\lim_{  \varepsilon\rightarrow0}\int_V \frac{(\dv\nv)^2}{\mathrm{vol}(V)} \dd V=\int_\esse \frac{(\dvs\nvs)^2}{{\mathrm{area}(\esse)}}\dd A,
\label{splay}
\ea
\ba
\lim_{  \varepsilon\rightarrow0}\int_V \frac{(\nv\cdot\rot\nv)^2}{\mathrm{vol}(V)}\dd V=\int_\esse \frac{(\nvs\cdot\rots\nvs)^2}{\mathrm{area}(\esse)}\dd A,
\label{twist}
\ea
\ba
\lim_{  \varepsilon \rightarrow0}\int_V\frac{|\nv\times\rot\nv|^2}{\mathrm{vol}(V)}\dd V=\int_\esse \frac{|\nvs\times\rots\nvs|^2}{\mathrm{area}(\esse)}\dd A.
\label{bend}
\ea
\end{proposition}

Therefore, from (\ref{null ss}), Proposition \ref{Frank theorem} and (\ref{limitvolume}), it follows that
\begin{align}
\nonumber
W_{OZF}&=\mathrm{vol}(V)\int_V \frac{K_1(\dv\nv)^2+K_2(\nv\cdot\rot\nv)^2+K_3|\nv\times\rot\nv|^2}{\mathrm{vol}(V)} \dd V\approx W_{OZF}^S 
\end{align}
for $  \varepsilon\ll1$, with $W_{OZF}^\esse$ as in (\ref{OZFshell}).

\section{ Geometrical identities}
Let us consider the orthogonal parameterization of $\esse$ introduced in Appendix A and set 
$$
x_1=u, \quad x_2=v, \quad \mathbf{g}_1=\phiv_{,u}=\sqrt{E}\ev_{1 s},\quad \mathbf{g}_2=\phiv_{,v}=\sqrt{G}\ev_{2 s}.
$$
The metric tensor  induced on $\esse$ by the Euclidean metric tensor,  written with respect to the system of local coordinates $(x_1,x_2)$, is
$$
g=E\dd x^1\otimes \dd x^1+G\dd x^2\ot \dd x^2.
$$
The Levi-Civita connection associated with the metric $g$ is defined by the Christoffel symbols
\begin{equation}
\left.\begin{array}{ll}
\displaystyle\Gamma_{11}^1=\frac{E_{,u}}{2E},\quad \Gamma_{11}^2=-\frac{E_{,v}}{2G}=\frac{E}{\sqrt{G}}\kappa_{1s}, \quad \Gamma_{12}^1=\Gamma_{21}^1=\frac{E_{,v}}{2E}=-\sqrt{G}\kappa_{1s},  \\
[5mm]
\displaystyle\Gamma_{12}^2=\Gamma_{21}^2=\frac{G_{,u}}{2G}=\sqrt{E}\kappa_{2s}, \quad \Gamma_{22}^1=-\frac{G_{,u}}{2E}=-\frac{G}{\sqrt{E}}\kappa_{2s}, \quad \Gamma_{22}^2=\frac{G_{,v}}{2G}.
\end{array}\right.
\end{equation}
Then, the $(0,4)$ curvature tensor of $\esse$ has components
\begin{align}
R_{\beta\gamma\delta\rho}&=g_{\rho\mu}\left(\frac{\partial \Gamma_{\gamma\delta}^\mu}{\partial x_\beta}-\frac{\partial \Gamma_{\beta\delta}^\mu}{\partial x_\gamma}+\Gamma_{\gamma\delta}^\lambda\Gamma_{\beta\lambda}^\mu-\Gamma_{\beta\delta}^\lambda\Gamma_{\gamma\lambda}^\mu\right)\\
\nonumber
&=EG(\grads\kappa_{2s}\cdot\ev_{1 s}-\grads \kappa_{1s}\cdot\ev_{2 s}+\kappa_{1s}^2+\kappa_{2s}^2)\epsilon_{\beta\gamma}\epsilon_{\delta\rho}\\
\nonumber
&=-EG\:(\bnus\cdot\rots\boldsymbol{\omega})\epsilon_{\beta\gamma}\epsilon_{\delta\rho}  \quad (\beta,\gamma,\delta,\rho,\mu,\lambda=1,2),
\end{align}
where  $\epsilon_{\beta\gamma}=\delta_{1\beta}\delta_{2\gamma}-\delta_{1\gamma}\delta_{2\beta}$ is the antisymmetric symbol and $
\boldsymbol{\omega}=-(\kappa_{1s}\ev_{1 s}+\kappa_{2s}\ev_{2 s})$ is the vector that parameterizes the spin connection $\Omega_{\beta\gamma\delta}$ (see \cite{giomi:2009}), that is
 $$
\Omega_{\beta\gamma\delta}=\ev_\gamma\cdot (D\ev_\delta)\ev_\beta=\omega_\beta\epsilon_{\gamma\delta} \quad (\beta,\gamma,\delta=1,2),
$$
where $D=\Pv\grads$ is the usual covariant derivative (see \cite{gurtin:1975}).
  It is well known that the Gaussian curvature of a surface equals the scalar curvature (see \cite{docarmo2}). Therefore
\begin{equation}\label{id}
\kg= \frac{1}{2}\sum_{\beta\neq\gamma} \frac{R_{\beta\gamma\beta\gamma}}{\det g}=-\bnus\cdot\rots\boldsymbol{\omega}.
\end{equation}

By means of (\ref{id}) we can prove  identity (\ref{ide}).
We first observe that $\kg \nvs + \kk \tavs = \grads \alpha - \bom$, by which $\kg \tavs - \kk \nvs = \bnus \times  (\grads \alpha - \bom)$, with $\alpha$ as in Section 3. Next, we recall the identity
\begin{align}\label{idc}
\bnus \cdot \rots(\grads f)= 0,
\end{align}
 that is valid for any smooth scalar field $f$ defined on $\esse$.
 Then, applying
the surface divergence theorem and  identities (\ref{ida1}), (\ref{id}) and (\ref{idc}) lead to
\begin{align}
\nonumber
\int_\esse &q (\grad_s q)\cdot (\kg \tav - \kk \nv) \dd A =  \frac{1}{2}\int_\esse \dvs[ q^2\bnus\times (\grads \alpha - \bom)] \dd A  \\
\nonumber
& - \frac{1}{2} \int_\esse q^2 \dvs[\bnus \times  (\grads \alpha - \bom)] \dd A\\
\nonumber
&=\frac{1}{2}\int_{\pt \esse}  q^2 [\bnus\times(\grads \alpha - \bom)]\cdot \kv \dd l+\frac{1}{2} \int_\esse q^2 \bnus \cdot\rots(\grads \alpha - \bom) \dd A\\
\nonumber
&=\frac{1}{2}\int_{\pt \esse}  q^2 (\grads \alpha - \bom)\cdot  \dd \mathbf{l}+\frac{1}{2} \int_\esse q^2 K \dd A,
\end{align}
where $\kv$ is the outward normal to the boundary $\partial\esse$ lying on the tangent plane.


\begin{thebibliography}{}

\bibitem[Bates, 2008]{bates:2008}
Bates, M.~A. (2008).
\newblock Nematic ordering and defects on the surface of a sphere: A monte
  carlo simulation study.
\newblock {\em J. Chem. Phys.}, 128(10):104707.

\bibitem[Biscari and Terentjev, 2006]{biscari:2006}
Biscari, P. and Terentjev, E.~M. (2006).
\newblock Nematic membranes: Shape instabilities of closed achiral vesicles.
\newblock {\em Phys. Rev. E}, 73(5):051706.

\bibitem[Bowick and Giomi, 2009]{giomi:2009}
Bowick, M. and Giomi, L. (2009).
\newblock Two-dimensional matter: order, curvature and defects.
\newblock {\em Advances in Physic}, 58(5):449--563.

\bibitem[Chen and Kamien, 2009]{kamien:2009}
Chen, B.~G. and Kamien, R.~D. (2009).
\newblock Nematic films and radially anisotropic delaunay surfaces.
\newblock {\em The European Physical Journal E: Soft Matter and Biological
  Physics}, 28(3):315-- 329.

\bibitem[de~Gennes and Prost, 1995]{degennes}
de~Gennes, P.-G. and Prost, J. (1995).
\newblock {\em The physics of liquid crystals}.
\newblock Oxford University Press.

\bibitem[do~Carmo, 1976]{docarmo}
do~Carmo, M.~P. (1976).
\newblock {\em Differential Geometry of Curves and Surfaces}.
\newblock Prentice-Hall, Englewood Cliffs, NJ.

\bibitem[do~Carmo, 1992]{docarmo2}
do~Carmo, M.~P. (1992).
\newblock {\em Riemannian Geometry}.
\newblock Birkh\"auser, Birkh\"auser Boston.

\bibitem[Fern\'andez-Nieves et~al., 2007]{Fernandez:2007}
Fern\'andez-Nieves, A., Vitelli, V., Utada, A.~S., Link, D.~R., M\'arquez, M.,
  Nelson, D.~R., and Weitz, D.~A. (2007).
\newblock Novel defect structures in nematic liquid crystal shells.
\newblock {\em Phys. Rev. Lett.}, 99(15).

\bibitem[Gurtin and Murdoch, 1975]{gurtin:1975}
Gurtin, M.~E. and Murdoch, A.~I. (1975).
\newblock A continuum theory of elastic material surfaces.
\newblock {\em Archive for Rational Mechanics and Analysis}, 57(4):291--323.

\bibitem[Helfrich and Prost, 1988]{helfrich:1988}
Helfrich, W. and Prost, J. (1988).
\newblock Intrinsic bending force in anisotropic membranes made of chiral
  molecules.
\newblock {\em Physical Review A}, 38(6).

\bibitem[Kralj et~al., 2011]{kralj:2011}
Kralj, S., Rosso, R., and Virga, E.~G. (2011).
\newblock Curvature control of valence on nematic shells.
\newblock {\em Soft Matter}, 7:670--683.

\bibitem[Longa et~al., 1987]{longa:1987}
Longa, L., Monselesan, D., and Trebin, H.-R. (1987).
\newblock An extension of the landau-ginzburg-de gennes theory for liquid
  crystals.
\newblock {\em Liq. Cryst.}, 2(6):769--796.

\bibitem[Lopez-Leon et~al., 2011a]{lopez-leon:2011prl}
Lopez-Leon, T., Fernandez-Nieves, A., Nobili, M., and Blanc, C. (2011a).
\newblock Nematic-smectic transition in spherical shells.
\newblock {\em Physical Review Letters}, 106(24):247802--.

\bibitem[Lopez-Leon et~al., 2011b]{lopez:2011}
Lopez-Leon, T., Koning, V., Devaiah, K. B.~S., Vitelli, V., and
  Fernandez-Nieves, A.~A. (2011b).
\newblock Frustrated nematic order in spherical geometries.
\newblock {\em Nat Phys}, 7(5):391--394.

\bibitem[Lubensky and MacKintosh, 1993]{lubensky:1993}
Lubensky, T.~C. and MacKintosh, F.~C. (1993).
\newblock Theory of \char16{}ripple\char17{} phases of lipid bilayers.
\newblock {\em Phys. Rev. Lett.}, 71(10):1565--1568.

\bibitem[Mbanga et~al., 2011]{santangelo:2011}
Mbanga, B.~L., Grason, G.~M., and Santangelo, C.~D. (2011).
\newblock Frustrated order on extrinsic geometries.
\newblock {\em arXiv:1108.1573v1}.

\bibitem[Napoli and Vergori, 2010]{napoli:2010}
Napoli, G. and Vergori, L. (2010).
\newblock Equilibrium of nematic vesicles.
\newblock {\em J. Phys. A: Math. Theor.}, 43(44):445207.

\bibitem[Nelson and Peliti, 1987]{nelson:1987}
Nelson, D. and Peliti, L. (1987).
\newblock Fluctuations in membranes with crystalline and hexatic order.
\newblock {\em J. Phys.}, 48(7):1085--1092.

\bibitem[Nelson, 2002]{nelson:2002}
Nelson, D.~R. (2002).
\newblock Toward a tetravalent chemistry of colloids.
\newblock {\em Nano Letters}, 2(10):1125--1129.

\bibitem[Rosso, 2003]{rosso:2003}
Rosso, R. (2003).
\newblock Curvature effects in vesicle-particle interactions.
\newblock {\em Proc. R. Soc. Lond. A}, 459(2032):829--852.

\bibitem[Selinger et~al., 2001]{selinger:2001}
Selinger, J.~V., Spector, M.~S., and Schnur, J.~M. (2001).
\newblock Theory of self-assembled tubules and helical ribbons.
\newblock {\em J Phys. Chem. B}, 105(30):7157--7169.

\bibitem[Shin et~al., 2008]{shin:2008}
Shin, H., Bowick, M.~J., and Xing, X. (2008).
\newblock Topological defects in spherical nematics.
\newblock {\em Phys. Rev. Lett.}, 101(3):037802.

\bibitem[Straley, 1971]{straley:1971}
Straley, J.~P. (1971).
\newblock Liquid crystals in two dimensions.
\newblock {\em Phy Rev A}, 4(2).

\bibitem[Tu and Ou-Yang, 2004]{tu:2004}
Tu, Z.~C. and Ou-Yang, Z.~C. (2004).
\newblock A geometric theory on the elasticity of bio-membranes.
\newblock {\em J. Phys. A: Math. Gen.}, 37(47):11407.

\bibitem[Tu and Seifert, 2007]{seifert:2007}
Tu, Z.~C. and Seifert, U. (2007).
\newblock Concise theory of chiral lipid membranes.
\newblock {\em Phys Rev. E}, 76(3 Pt 1):031603.

\bibitem[Turzi, 2007]{Turzi}
Turzi, S. (2007).
\newblock {\em Distortion-induced effects in nematic liquid crystals}.
\newblock PhD Thesis, Politecnico di Milano.

\bibitem[Virga, 1994]{virga}
Virga, E.~G. (1994).
\newblock {\em Variational Theories For Liquid Crystals Variational Theories
  For Liquid Crystals}.
\newblock Chapman-Hall, London.

\bibitem[Vitelli and Nelson, 2006]{vitelli:2006}
Vitelli, V. and Nelson, D.~R. (2006).
\newblock Nematic textures in spherical shells.
\newblock {\em Phys. Rev. E}, 74(2):021711.

\end{thebibliography}
\end{document}